\documentclass[aps,prb,twocolumn,amsmath,showpacs]{revtex4} 

\usepackage{bm}
\usepackage{graphicx}

\begin{document}

\title{Analytical investigation of magnetic field distributions around superconducting strips on ferromagnetic substrates 
}
\author{Yasunori Mawatari}
\affiliation{%
	National Institute of Advanced Industrial Science and Technology (AIST), 
	Tsukuba, Ibaraki 305--8568, Japan}

\date{October 11, 2007}

\begin{abstract}
The complex-field approach is developed to derive analytical expressions of the magnetic field distributions around superconducting strips on ferromagnetic substrates (SC/FM strips). 
We consider the ferromagnetic substrates as ideal soft magnets with an infinite magnetic permeability, neglecting the ferromagnetic hysteresis. 
On the basis of the critical state model for a superconducting strip, the ac susceptibility $\chi_1'+i\chi_1''$ of a SC/FM strip exposed to a perpendicular ac magnetic field is theoretically investigated, and the results are compared with those for superconducting strips on nonmagnetic substrates (SC/NM strips). 
The real part $\chi_1'$ for $H_0/j_cd_s\to 0$ (where $H_0$ is the amplitude of the ac magnetic field, $j_c$ is the critical current density, and $d_s$ is the thickness of the superconducting strip) of a SC/FM strip is $3/4$ of that of a SC/NM strip. 
The imaginary part $\chi_1''$ (or ac loss $Q$) for $H_0/j_cd_s<0.14$ of a SC/FM strip is larger than that of a SC/NM strip, even when the ferromagnetic hysteresis is neglected, and this enhancement of $\chi_1''$ (or $Q$) is due to the edge effect of the ferromagnetic substrate. 
\end{abstract}
\maketitle

\section{Introduction} 
High-temperature superconducting coated conductors have been developed for applications in electric power devices, and remarkable progress toward high-current and long-length conductors has recently been reported.~\cite{Shiohara07} 
In superconducting coated conductors, the superconducting layers are generally fabricated on metallic substrates with oxide buffer layers, and ferromagnetic materials (e.g., Ni alloys) are promising candidates for the metallic substrates.~\cite{Goyal96,Ijaduola04} 
The magnetic behavior of ferromagnetic substrates can strongly affect the electromagnetic response of superconducting coated conductors. 
Although the effects of ferromagnetic substrates on ac losses of superconducting coated conductors have been extensively investigated experimentally and numerically,~\cite{Duckworth03,Tsukamoto05,Gomory06a,Suenaga06,Claassen07,Nguyen07,Tsukamoto07,Amemiya07,Miyagi07,Suenaga} ac losses in superconducting strips on ferromagnetic substrates have not been investigated analytically. 
Genenko {\it et al}.~\cite{Genenko99} analytically investigated the magnetic-field and current distributions in superconducting strips surrounded by soft magnets, but they did not consider realistic geometries similar to those of coated conductors. 

In the present paper, we develop a theoretical framework to investigate electromagnetic response of superconducting strips on ferromagnetic substrates (SC/FM strips), and we compare the results with those for superconducting strips on nonmagnetic substrates (SC/NM strips).~\cite{Halse70,Brandt93,Zeldov94a} 
Section~\ref{sec_model} introduces the theoretical models and methods that we used to investigate the magnetic field around SC/FM strips. 
Section~\ref{sec_Meissner} gives the theoretical results for the magnetic field distribution around a SC/FM strip in which the superconducting strip is in the ideal Meissner state and exposed either to a perpendicular magnetic field, a parallel magnetic field, or a transport current.
Section~\ref{sec_critical-state} gives the theoretical results for dc and ac response of a SC/FM strip in which the superconducting strip is in the critical state and the SC/FM strip is exposed to a perpendicular magnetic field. 
Section~\ref{sec_summary} describes the comparison of our theoretical results with experimental data by Suenaga {\it et al}.~\cite{Suenaga} and summarizes our results.

\section{\label{sec_model}%
Model} 
In this section, the configuration of a SC/FM strip is defined, and the basic theoretical models used to investigate the electromagnetic response of a SC/FM strip are introduced. 

\begin{figure}[b]
	\includegraphics{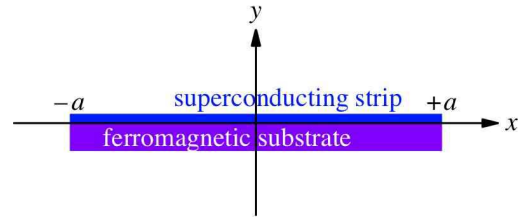}
\caption{(Color onine) %
Cross section of a SC/FM strip in the $xy$ plane. 
Superconducting strip is situated at $|x|<a$ and $0<y<d_s$, and ferromagnetic substrate at $|x|<a$ and $-d_m<y<0$, where $d_s+d_m\ll 2a$. 
}
\label{Fig_SC/FM-strip}
\end{figure}

Consider a SC/FM strip of width $2a$, total thickness $d_s+d_m$, and infinite length along the $z$ axis, as shown in Fig.~\ref{Fig_SC/FM-strip}. 
This strip consists of a superconducting strip whose thickness is $d_s$, and a ferromagnetic substrate whose thickness is $d_m$, where $d_s+d_m\ll 2a$. 
Let $\epsilon\equiv \max(d_s,d_m)$ be a positive infinitesimal, and the thin strip limit of $\epsilon\to +0$ enables analytical expressions of the magnetic field distribution around a SC/FM strip to be derived as follows.

\subsection{Complex field} 
To analyze a two-dimensional magnetic field, $\bm{H}= H_x(x,y)\hat{\bm{x}} +H_y(x,y)\hat{\bm{y}}$, we consider the complex field~\cite{Beth66,Clem73,Zeldov94b,Mawatari01,Brojeny02,Sonin02,Mawatari03,Mawatari06,Clem06} 
\begin{equation}
	{\cal H}(\zeta)= H_y(x,y)+iH_x(x,y) , 
\label{eq_complex-field}
\end{equation}
which is the analytical function of the complex variable $\zeta=x+iy$ outside of a SC/FM strip. 
Applying Cauchy's integral formula~\cite{Arfken-Weber} to ${\cal H}(\zeta)$ yields the following: 
\begin{eqnarray}
	{\cal H}(\zeta) &=& \frac{1}{2\pi i}\oint_C d\zeta' 
		\frac{{\cal H}(\zeta')}{\zeta'-\zeta} 
\nonumber\\
	&=& \lim_{R\to\infty} \frac{1}{2\pi} \int_0^{2\pi} d\theta\, 
		\frac{Re^{i\theta}\, {\cal H}(Re^{i\theta})}{Re^{i\theta} -\zeta} 
\nonumber\\
	&& {}+\frac{1}{2\pi i}\int_{-\infty}^{+\infty} dx' 
		\frac{{\cal H}(x'+i\epsilon)-{\cal H}(x'-i\epsilon)}{x'-\zeta} , 
	\quad
\label{eq_Cauchy}
\end{eqnarray}
where the closed contour $C$ has three components: a line just above the real axis (i.e., $\zeta'=x'+i\epsilon$ from $x'=-\infty$ to $x'=+\infty$), an infinite circle (i.e., $\zeta'=Re^{i\theta}$ from $\theta=0$ to $\theta=2\pi$ with $R\to\infty$), and a line just below the real axis (i.e., $\zeta'=x'-i\epsilon$ from $x'=+\infty$ to $x'=-\infty$); see Fig.~7 in Ref.~\onlinecite{Mawatari06}. 
Substitution of ${\cal H}(\zeta)\to H_{ay}+iH_{ax}$ for $|\zeta|\to\infty$ and ${\cal H}(x'+i\epsilon)={\cal H}(x'-i\epsilon)$ for $|x'|>a$ into Eq.~\eqref{eq_Cauchy} leads to the generalized Biot-Savart law for a SC/FM strip: 
\begin{equation}
	{\cal H}(\zeta)= (H_{ay}+iH_{ax}) +\frac{1}{2\pi}\int_{-a}^{+a} dx' 
		\frac{K_z(x') +i\sigma_m(x')}{\zeta-x'} , 
\label{eq_Cauchy-Biot-Savart}
\end{equation}
where $\bm{H}_a= H_{ax}\hat{\bm{x}} +H_{ay}\hat{\bm{y}}$ is a uniform applied magnetic field, $K_z(x)$ is the sheet current in a superconducting strip, and $\sigma_m(x)$ is the effective sheet magnetic charge~\cite{Jackson} in a ferromagnetic substrate. 
The $K_z(x)$ and $\sigma_m(x)$ are defined by 
\begin{eqnarray}
	K_z(x) &=&  H_x(x,-\epsilon) -H_x(x,+\epsilon) , 
\label{eq_sheet-current}\\
	\sigma_m(x) &=& H_y(x,+\epsilon) -H_y(x,-\epsilon) , 
\label{eq_magnetic-charge}
\end{eqnarray}
respectively. 
The net magnetic charge is zero; that is, $\int_{-a}^{+a} \sigma_m(x)dx=0$. 

The multipole expansion of Eq.~\eqref{eq_Cauchy-Biot-Savart} for $|\zeta|/a\to\infty$ is given by~\cite{Mawatari03} 
\begin{equation}
	{\cal H}(\zeta)\to (H_{ay}+iH_{ax}) +\frac{I_z}{2\pi \zeta}
		+ \frac{-m_y+im_x}{2\pi\zeta^2} +\cdots , 
\label{eq_H-infinity}
\end{equation}
where $I_z=\int_{-a}^{+a} K_z(x)dx$ is the transport current flowing in a superconducting strip. 
The $\bm{m}= m_x\hat{\bm{x}} +m_y\hat{\bm{y}}$ is the magnetic moment per unit length of a SC/FM strip: 
\begin{eqnarray}
	m_y &=& -\int_{-a}^{+a}dx\, x K_z(x) 
\nonumber\\
	&=& \int_{-a}^{+a}dx\, x \left[ H_x(x,+\epsilon) -H_x(x,-\epsilon) \right] , 
\label{eq_my-Kz}\\
	m_x &=& \int_{-a}^{+a}dx\, x \sigma_m(x) 
\nonumber\\
	&=& \int_{-a}^{+a}dx\, x \left[ H_y(x,+\epsilon) -H_y(x,-\epsilon) \right] . 
\label{eq_mx-sigma}
\end{eqnarray}
The $m_y$ is induced by $K_z(x)$ in a superconducting strip, whereas $m_x$ is induced by $\sigma_m(x)$ in a ferromagnetic substrate. 

The complex potential defined by 
\begin{equation}
	{\cal G}(\zeta)= \int{\cal H}(\zeta)d\zeta 
\label{eq_G-int-H}
\end{equation}
is convenient for visualizing the magnetic field lines around a SC/FM strip, because the contour lines of the real part of Eq.~\eqref{eq_G-int-H}, ${\rm Re}\,{\cal G}(\zeta)$, correspond to the magnetic field lines.~\cite{Mawatari06,Clem06} 

The variable transformation (i.e., the conformal mapping) defined by 
\begin{equation}
	\eta= i\sqrt{\zeta^2-a^2} , \quad
	\zeta=-i\sqrt{\eta^2-a^2} 
\label{eq_eta-zeta}
\end{equation}
is useful for analyzing ${\cal H}(\zeta)$ for a SC/FM strip.

\subsection{\label{sec_FM-substrate}%
Ferromagnetic substrate} 
The ferromagnetic materials (e.g., Ni-W alloys) used as metal substrates for coated conductors are classified as soft magnets.~\cite{Ijaduola04,Suenaga06} 
For simplicity, in the present paper we consider the ferromagnetic substrates as {\em ideal soft magnets}, in which the relationship between the magnetic induction $\bm{B}$ and the magnetic field $\bm{H}$ is given by~\cite{Genenko99} $\bm{B}=\mu_m\bm{H}$, where $\mu_m$ is much larger than the magnetic permeability of the vacuum $\mu_0$ (i.e., $\mu_m/\mu_0\gg 1$). 
Also, in the ideal soft magnet model, we neglect the ferromagnetic hysteresis and assume that the saturation field $H_s$ is much larger than $|\bm{H}|$. 

In the infinite permeability limit ($\mu_m/\mu_0\to\infty$), the $\bm{H}=\bm{B}/\mu_0$ outside of the ideal soft magnet has only a perpendicular component at the surface.~\cite{Jackson} 
Therefore, the boundary condition at the surface of a ferromagnetic substrate (i.e., at $y=-\epsilon$) is given by 
\begin{equation}
	H_x(x,-\epsilon)={\rm Im}\,{\cal H}(x-i\epsilon)=0 
	\quad\mbox{for } |x|<a . 
\label{eq_Hx(x,0)=0}
\end{equation}
The magnetic field distribution for a large but finite permeability ($\mu_m/\mu_0\gg 1$) is not significantly different from that for an infinite permeability ($\mu_m/\mu_0\to\infty$).~\cite{Genenko99} 
The simple boundary condition of Eq.~\eqref{eq_Hx(x,0)=0} thus enables analytical expressions of ${\cal H}(\zeta)$ to be derived. 

\subsection{Superconducting strip} 
The ideal Meissner state model (Sec.~\ref{sec_Meissner}) and the critical state model (Sec.~\ref{sec_critical-state}) are adopted to investigate the electromagnetic response of a superconducting strip in a SC/FM strip. 

In Sec.~\ref{sec_Meissner}, the ideal case is considered, namely, when a superconducting strip is in the ideal Meissner state. 
In this state the magnetic flux does not penetrate a superconducting strip, and consequently, the perpendicular component of the magnetic field vanishes, that is, $H_y(x,+\epsilon)=0$. 

In Sec.~\ref{sec_critical-state}, a more realistic case is considered, and thus we used the critical state model with constant critical current density $j_c$, as in the Bean model.~\cite{Bean62} 
Similar to earlier calculations,~\cite{Halse70,Brandt93,Zeldov94a} the effects of the lower critical field $H_{c1}$ are neglected (i.e., $H_{c1}\ll |\bm{H}|$), and thus the $\bm{B}$-$\bm{H}$ relationship is simply given by $\bm{B}= \mu_0\bm{H}$. 
In the critical state model, the magnetic flux penetrates and $K_z$ reaches its critical value $j_cd_s$ near the edges of the superconducting strip. 
The $|K_z(x)|= j_cd_s$ holds in the flux-filled region [where $H_y(x,+\epsilon)\neq 0$] near the edges in a superconducting strip, whereas the flux-free region [where $H_y(x,+\epsilon)=0$] exists near the center of a superconducting strip.

\section{\label{sec_Meissner}%
Ideal Meissner state} 
In this section, we consider the complex field ${\cal H}(\zeta)$ for a SC/FM strip in which the superconducting strip is in the ideal Meissner state. 
In the ideal Meissner state, the magnetic field component perpendicular to the surface of the superconducting strip at $y=+\epsilon$ is zero: 
\begin{equation}
	H_y(x,+\epsilon)={\rm Re}\,{\cal H}(x+i\epsilon)=0 
	\quad\mbox{for } |x|<a . 
\label{eq_Hy(x,0)=0}
\end{equation}
In addition to the boundary conditions given by Eqs.~\eqref{eq_Hx(x,0)=0} and \eqref{eq_Hy(x,0)=0}, further conditions depending on $H_{ax}$, $H_{ay}$, and $I_z$ are needed to determine ${\cal H}(\zeta)$, as shown in the following subsections.

\subsection{Response to a perpendicular magnetic field} 
Here, the ${\cal H}(\zeta)$ for a SC/FM strip is presented for the case when a SC/FM strip is exposed to a perpendicular magnetic field $\bm{H}_a=H_{ay}\hat{\bm{y}}$; that is, when $H_{ay}\neq 0$ and $H_{ax}=I_z=0$.

The corresponding complex field for $H_{ax}=I_z=0$, which satisfies the conditions given in Eqs.~\eqref{eq_Hx(x,0)=0} and \eqref{eq_Hy(x,0)=0}, is expressed as
\begin{equation}
	{\cal H}(\zeta)= H_{ay} \left(1-\frac{a}{2\eta}\right) 
		\sqrt{\frac{\eta+a}{\eta}} , 
\label{eq_Hay-Meissner}
\end{equation}
where $\eta=\eta(\zeta)$ is a function of $\zeta$ given by Eq.~\eqref{eq_eta-zeta}. 
Equation~\eqref{eq_Hay-Meissner} is expanded for $|\zeta|\sim |\eta|\to\infty$ as 
\begin{equation}
	{\cal H}(\zeta)\to H_{ay}\left( 1+\frac{3a^2}{8\zeta^2} +\cdots \right) . 
\label{eq_Hay-Meissner-infinity}
\end{equation}
Comparison of Eqs.~\eqref{eq_H-infinity} and \eqref{eq_Hay-Meissner-infinity} yields the magnetic moment per unit length, 
\begin{equation}
	m_y = \chi_{0y} H_{ay} , 
\label{eq_my-Meissner}
\end{equation}
where $\chi_{0y}$ is the magnetic susceptibility given by 
\begin{equation}
	\chi_{0y} = -(3\pi/4) a^2 .
\label{eq_X0-Hay-Meissner}
\end{equation}
Here we defined $\chi_{0y}$ as the ratio of $m_y$ in units of ($\rm A\cdot m$) to $H_{ay}$ in units of ($\rm A/m$), such that $\chi_{0y}$ is in units of ($\rm m^2$). 
Equation~\eqref{eq_X0-Hay-Meissner} corresponds to $3/4$ of the magnetic  susceptibility of a SC/NM strip in the ideal Meissner state,~\cite{Brandt93} $\chi_{0y}=-\pi a^2$. 

\begin{figure*}[bt]
	\includegraphics{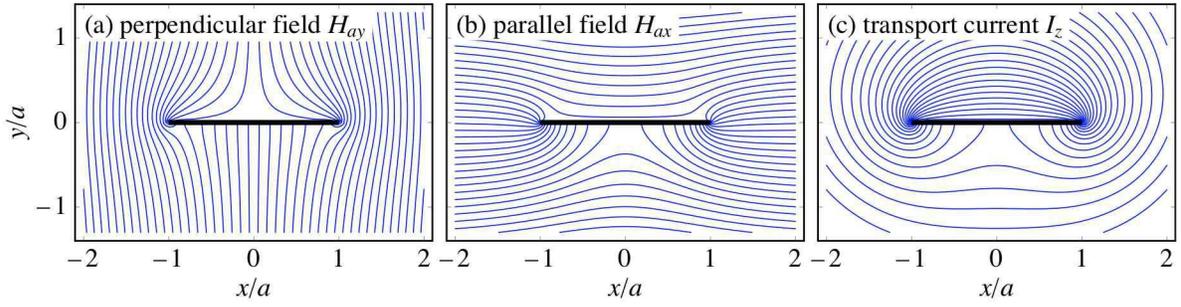}
\caption{(Color onine) %
Magnetic field lines [i.e., contour lines of ${\rm Re}\,{\cal G}(\zeta)$] around a SC/FM strip in which the superconducting strip is in the ideal Meissner state: 
(a) in a perpendicular magnetic field $H_{ay}$, (b) in a parallel magnetic field $H_{ax}$, and (c) with a transport current $I_z$. 
Thick horizontal bar at $-1<x/a<1$ and $y=0$ denotes the SC/FM strip. 
}
\label{Fig_fls-Meissner}
\end{figure*}

The complex potential calculated by substituting Eq.~\eqref{eq_Hay-Meissner} into Eq.~\eqref{eq_G-int-H} is given by 
\begin{equation}
	{\cal G}(\zeta)= -iH_{ay}\sqrt{\eta(\eta-a)} . 
\label{eq_G-Hay-Meissner}
\end{equation}
Figure~\ref{Fig_fls-Meissner}(a) shows the magnetic field lines calculated from ${\rm Re}\,{\cal G}(\zeta)$ with Eqs.~\eqref{eq_eta-zeta} and \eqref{eq_G-Hay-Meissner}. 
At the surface of the ferromagnetic substrate at $y=-\epsilon$, the $\bm{H}$ near the edges is {\em downward} [i.e., $H_y(x,-\epsilon)<0$ when $H_{ay}>0$] for $x_p<|x|<a$, whereas $\bm{H}$ is mostly {\em upward} [i.e., $H_y(x,-\epsilon)>0$ when $H_{ay}>0$] for $|x|<x_p$, where $x_p/a= \sqrt{3}/2=0.866$. 
Such pronounced behavior in $\bm{H}$ due to a ferromagnetic substrate results in the concentration of the magnetic field near the edges of the SC/FM strip. 

\subsection{Response to a parallel magnetic field} 
Here, the ${\cal H}(\zeta)$ for a SC/FM strip is presented for the case when a SC/FM strip is exposed to a parallel magnetic field $\bm{H}_a=H_{ax}\hat{\bm{x}}$; that is, when $H_{ax}\neq 0$ and $H_{ay}=I_z=0$.

The corresponding complex field for $H_{ay}=I_z=0$, which satisfies the conditions given in Eqs.~\eqref{eq_Hx(x,0)=0} and \eqref{eq_Hy(x,0)=0}, is expressed as 
\begin{equation}
	{\cal H}(\zeta)= iH_{ax} \left(1+\frac{a}{2\eta}\right) 
		\sqrt{\frac{\eta-a}{\eta}} , 
\label{eq_Hax-Meissner}
\end{equation}
where $\eta$ is a function of $\zeta$ by Eq.~\eqref{eq_eta-zeta}. 
Equation~\eqref{eq_Hax-Meissner} is expanded for $|\zeta|\sim |\eta|\to\infty$ as 
\begin{equation}
	{\cal H}(\zeta)\to iH_{ax}\left( 1+\frac{3a^2}{8\zeta^2} +\cdots \right) . 
\label{eq_Hax-Meissner-infinity}
\end{equation}
Comparison of Eqs.~\eqref{eq_H-infinity} and \eqref{eq_Hax-Meissner-infinity} yields the magnetic moment per unit length, 
\begin{equation}
	m_x = \chi_{0x} H_{ax} , 
\label{eq_mx-Meissner}
\end{equation}
where $\chi_{0x}$ is the magnetic susceptibility given by 
\begin{equation}
	\chi_{0x} = +(3\pi/4) a^2 . 
\label{eq_X0-Hax-Meissner}
\end{equation}
Equation~\eqref{eq_X0-Hax-Meissner} corresponds to $3/4$ of the magnetic  susceptibility of a ferromagnetic strip without a superconducting strip, $\chi_{0x}= +\pi a^2$.

The complex potential calculated by substituting Eq.~\eqref{eq_Hax-Meissner} into Eq.~\eqref{eq_G-int-H} is given by 
\begin{equation}
	{\cal G}(\zeta)= H_{ax}\sqrt{\eta(\eta+a)} . 
\label{eq_G-Hax-Meissner}
\end{equation}
Figure~\ref{Fig_fls-Meissner}(b) shows the magnetic field lines calculated from Eqs.~\eqref{eq_eta-zeta} and \eqref{eq_G-Hax-Meissner}. 
At the surface of the superconducting strip at $y=+\epsilon$, the $\bm{H}$ near the edges is {\em leftward} [i.e., $H_x(x,+\epsilon)<0$ when $H_{ax}>0$] for $x_p<|x|<a$, whereas $\bm{H}$ is mostly {\em rightward} [i.e., $H_x(x,+\epsilon)>0$ when $H_{ax}>0$] for $|x|<x_p$, where $x_p/a= \sqrt{3}/2=0.866$.

\subsection{Response to a transport current} 
Here, the ${\cal H}(\zeta)$ for a SC/FM strip is presented for the case when a superconducting strip in a SC/FM strip carries a transport current $I_z$; that is, when $I_z\neq 0$ and $H_{ax}=H_{ay}=0$. 

The corresponding complex field for $H_{ax}=H_{ay}=0$, which satisfies the boundary conditions given in Eqs.~\eqref{eq_Hx(x,0)=0} and \eqref{eq_Hy(x,0)=0}, is expressed as 
\begin{equation}
	{\cal H}(\zeta)= i\frac{I_z}{2\pi} 
		\frac{1}{\eta} \sqrt{\frac{\eta-a}{\eta}} , 
\label{eq_Iz-Meissner}
\end{equation}
where $\eta$ is a function of $\zeta$ by Eq.~\eqref{eq_eta-zeta}. 
Equation~\eqref{eq_Iz-Meissner} is expanded for $|\zeta|\sim |\eta|\to\infty$ as 
\begin{equation}
	{\cal H}(\zeta)\to \frac{I_z}{2\pi} 
		\left( \frac{1}{\zeta} +\frac{ia}{2\zeta^2} +\cdots \right) . 
\label{eq_Iz-Meissner-infinity}
\end{equation}
Comparison of Eqs.~\eqref{eq_H-infinity} and \eqref{eq_Iz-Meissner-infinity} reveals that, despite $H_{ax}=0$, the magnetic moment per unit length, 
\begin{equation}
	m_x= aI_z/2 , 
\label{eq_mx-Iz-Meissner}
\end{equation}
is induced from $I_z$. 

The complex potential calculated by substituting Eq.~\eqref{eq_Iz-Meissner} into Eq.~\eqref{eq_G-int-H} is given by 
\begin{equation}
	{\cal G}(\zeta)= (I_z/\pi)\, 
		\mbox{arcsinh} \left(\sqrt{\eta/a}\right) . 
\label{eq_G-Iz-Meissner}
\end{equation}
Figure~\ref{Fig_fls-Meissner}(c) shows the magnetic field lines calculated from Eqs.~\eqref{eq_eta-zeta} and \eqref{eq_G-Iz-Meissner}.

\section{\label{sec_critical-state}%
Critical state} 
In this section, we consider ${\cal H}(\zeta)$ for a SC/FM strip in which the superconducting strip is in the critical state. 
A SC/FM strip is exposed to a perpendicular magnetic field $\bm{H}_a=H_{ay}\hat{\bm{y}}$ and carries no net transport current (i.e., $H_{ax}=I_z=0$), where $H_{ay}$ is either dc or ac magnetic field. 

\subsection{Response to a dc magnetic field} 
Here, we consider the case when a SC/FM strip is exposed to a dc magnetic field $H_{ay}=H_0$, which is fixed after monotonically increased from $H_{ay}=0$. 

The magnetic flux penetrates the superconducting strip near the edges, and the sheet current density $K_x(x)=-H_x(x,+\epsilon)$ [from Eqs.~\eqref{eq_sheet-current} and \eqref{eq_Hx(x,0)=0}] reaches its critical value, 
\begin{equation}
	H_x(x,+\epsilon)={\rm Im}\,{\cal H}(x+i\epsilon)= -{\rm sgn}(x)j_cd_s 
	\ \ \mbox{for } b_0<|x|<a . 
\label{eq_Hx(x,0)=jcd}
\end{equation}
In contrast, the magnetic flux does not penetrate the inner region, 
\begin{equation}
	H_y(x,+\epsilon)={\rm Re}\,{\cal H}(x+i\epsilon)=0 
	\quad\mbox{for } |x|<b_0 , 
\label{eq_Hy(x,0)=0_b0}
\end{equation}
where $b_0$ is the parameter for the flux front. 
At the surface of the ferromagnetic substrate, the parallel component of the magnetic field is zero, as required by the boundary condition in Eq.~\eqref{eq_Hx(x,0)=0}. 
The corresponding complex field that satisfies the conditions given in Eqs.~\eqref{eq_Hx(x,0)=0}, \eqref{eq_Hx(x,0)=jcd}, and \eqref{eq_Hy(x,0)=0_b0} is expressed by 
\begin{equation}
	\frac{{\cal H}(\zeta)}{2j_cd_s/\pi} = 
		{\rm arctanh}\left[ \sqrt{%
		\frac{\beta_0(\eta+a)}{a(\eta+\beta_0)} } \right] 
		-\frac{\sqrt{a\beta_0(\eta+a)(\eta+\beta_0)}}{%
		(a+\beta_0)\eta} , 
\label{eq_H-critical}
\end{equation}
where $\eta$ is given by Eq.~\eqref{eq_eta-zeta}, and $\beta_0$ is given by 
\begin{equation}
	\beta_0= \sqrt{a^2-b_0^2} . 
\label{b0-beta0}
\end{equation}

\begin{figure}[bt]
	\includegraphics{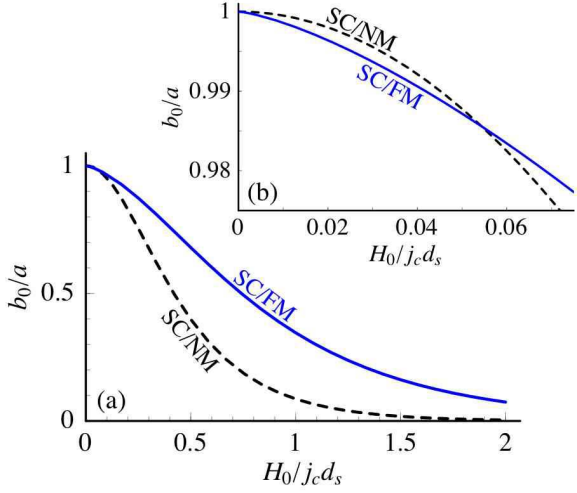}
\caption{(Color onine) %
Parameter for the flux front $b_0$ (in units of $a$) as a function of an applied magnetic field $H_0$ (in units of $j_cd_s$) (a) for $0<H_0/j_cd_s<2$ and (b) for $0<H_0/j_cd_s<0.075$. 
Solid lines represent $b_0$ vs $H_0$ for a SC/FM strip obtained from Eqs.~\eqref{b0-beta0} and \eqref{eq_H0-beta0}, and dashed lines represent $b_0$ vs $H_0$ for a SC/NM strip.~\cite{Halse70,Brandt93,Zeldov94a} 
}
\label{Fig_flux-front}
\end{figure}

Equation~\eqref{eq_H-critical} is expanded for $|\zeta|\sim |\eta|\to\infty$ as 
\begin{equation}
	\frac{{\cal H}(\zeta)}{2j_cd_s/\pi} \to 
		{\rm arctanh}\left( \sqrt{\frac{\beta_0}{a}} \right) 
		-\frac{\sqrt{a\beta_0}}{a+\beta_0} 
		+ \frac{(a\beta_0)^{3/2}}{2(a+\beta_0) \zeta^2} 
		+\cdots . 
\label{eq_H-critical-infinity}
\end{equation}
Comparison of Eqs.~\eqref{eq_H-infinity} and \eqref{eq_H-critical-infinity} yields the following relationship between $H_0$ and $\beta_0$: 
\begin{equation}
	\frac{H_0}{2j_cd_s/\pi} = 
		{\rm arctanh}\left( \sqrt{\frac{\beta_0}{a}} \right) 
		-\frac{\sqrt{a\beta_0}}{a+\beta_0} . 
\label{eq_H0-beta0}
\end{equation}
The parameter for the flux front $b_0$ is obtained as a function of the applied magnetic field $H_0$ by eliminating $\beta_0$ from Eqs.~\eqref{b0-beta0} and \eqref{eq_H0-beta0}. 
The resulting $b_0$ vs $H_0$ for a SC/FM strip is shown as the solid lines in Fig.~\ref{Fig_flux-front}, and $b_0=a/\cosh(\pi H_0/j_cd_s)$ for a SC/NM strip~\cite{Halse70,Brandt93,Zeldov94a} is shown as the dashed lines. 
When $H_0/j_cd_s> 0.054$, the magnetic flux penetration into a SC/FM strip is {\em slower} than that into a SC/NM strip (i.e., $b_0$ for a SC/FM strip is larger than $b_0$ for a SC/NM strip), as seen in Fig.~\ref{Fig_flux-front}(a); whereas when $H_0/j_cd_s< 0.054$, the penetration is {\em faster} into a SC/FM strip, as seen in Fig.~\ref{Fig_flux-front}(b). 

\begin{figure}[bt]
	\includegraphics{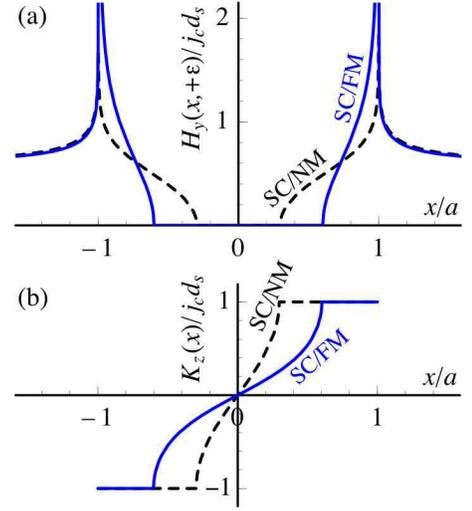}
\caption{(Color onine) %
Distributions of (a) perpendicular magnetic field $H_y(x,+\epsilon)$ (in units of $j_cd_s$) and (b) sheet current $K_z(x)$ (in units of $j_cd_s$) as a function of $x$ (in units of $a$) for $H_0/j_cd_s=0.6$. 
Solid lines show $H_y={\rm Re}\,{\cal H}(x+i\epsilon)$ and $K_z={\rm Im}\,[{\cal H}(x-i\epsilon)-{\cal H}(x+i\epsilon)]$ calculated from Eq.~\eqref{eq_H-critical} for a SC/FM strip ($b_0/a=0.602$), and dashed lines show $H_y$ and $K_z$ for a SC/NM strip~\cite{Halse70,Brandt93,Zeldov94a} ($b_0/a=0.297$). 
}
\label{Fig_Hy-Kz-x}
\end{figure}

The solid lines in Fig.~\ref{Fig_Hy-Kz-x} show the calculated distributions of $H_y(x,+\epsilon)$ and $K_z(x)$ for a SC/FM strip, and the dashed lines show $H_y$ and $K_z$ for a SC/NM strip.~\cite{Halse70,Brandt93,Zeldov94a} 
The $|H_y|$ and $|K_z|$ of a SC/FM strip are {\em smaller} than those of a SC/NM strip in the inner region ($|x|/a\lesssim 0.6$), whereas $|H_y|$ of a SC/FM strip is {\em larger} in the outer region ($0.7\lesssim |x|/a<1$). 
The stronger magnetic field near the edges of a SC/FM strip is due to the edge effect of the ferromagnetic substrate, and is responsible for the faster magnetic-flux penetration (i.e., smaller $b_0$) into a SC/FM strip in the weak magnetic field regime ($H_0/j_cd_s< 0.054$), as shown in Fig.~\ref{Fig_flux-front}(b). 

From Eqs.~\eqref{eq_H-infinity} and \eqref{eq_H-critical-infinity}, the magnetic moment per unit length is given by $m_y=m_0(H_0)$, where 
\begin{equation}
	m_0 = -2j_cd_s \frac{(a\beta_0)^{3/2}}{a+\beta_0} . 
\label{eq_m0-critical}
\end{equation}
The $m_0$ is obtained as a function of $H_0$ by eliminating $\beta_0$ in Eqs.~\eqref{eq_H0-beta0} and \eqref{eq_m0-critical}, and the differential susceptibility is expressed as 
\begin{equation}
	\frac{\partial m_0}{\partial H_0} 
	= \frac{\partial m_0/\partial \beta_0}{\partial H_0/\partial \beta_0} 
	= -\frac{\pi}{4} (3a+\beta_0)(a-\beta_0) . 
\label{eq_dm0/dH0}
\end{equation}
For a weak magnetic field of $H_0\ll j_cd_s$ (i.e., $\beta_0\ll a$), Eq.~\eqref{eq_dm0/dH0} is reduced to $\partial m_0/\partial H_0\simeq -(3\pi/4)a^2$, which corresponds to Eq.~\eqref{eq_X0-Hay-Meissner}.

\subsection{Response to an ac magnetic field} 
Here, we consider the case when a SC/FM strip is exposed to an ac magnetic field $H_{ay}=H_0\cos\omega t$. 

The magnetic moment per unit length $m_y(t)$ for an ac magnetic field is expressed as~\cite{Halse70,Brandt97} 
\begin{equation}
	m_y(t) = +m_0(H_0) -2m_0\bigl(H_0(1-\cos\omega t)/2\bigr) 
\label{eq_my-H-dec}
\end{equation}
for $0<\omega t<\pi$, and 
\begin{equation}
	m_y(t) = -m_0(H_0) +2m_0\bigl(H_0(1+\cos\omega t)/2\bigr) 
\label{eq_my-H-inc}
\end{equation}
for $\pi<\omega t<2\pi$, where $m_0(H_0)$ is given by Eqs.~\eqref{eq_H0-beta0} and \eqref{eq_m0-critical} by eliminating $\beta_0$. 
The $m_y(t)$ can be expressed as the Fourier series: 
\begin{eqnarray}
	m_y(t) &=& H_0\sum_{n=1}^{\infty} 
		\left( \chi_n'\cos\omega t +\chi_n''\sin\omega t \right) 
\nonumber\\
	&=& H_0\sum_{n=1}^{\infty} {\rm Re}\,\left[%
		\left( \chi_n'+i\chi_n'' \right) e^{-in\omega t}\right] , 
\label{eq_my-Fourier}
\end{eqnarray}
where the ac susceptibility $\chi_n'+i\chi_n''$ is calculated as~\cite{Clem94} 
\begin{equation}
	\chi_n'+i\chi_n'' = 
		\frac{1}{\pi H_0} \int_0^{2\pi}d(\omega t) 
		m_y(t) e^{in\omega t} . 
\label{eq_chi-calc}
\end{equation}
Substitution of Eqs.~\eqref{eq_my-H-dec} and \eqref{eq_my-H-inc} into Eq.~\eqref{eq_chi-calc} yields $\chi_n'=\chi_n''=0$ for even $n$. 
The ac susceptibility for odd $n$ is given by 
\begin{eqnarray}
	\lefteqn{ \chi_n'+i\chi_n'' } 
\nonumber\\
	&=& 
		\frac{2}{\pi H_0} \int_0^{\pi}d\theta\, e^{in\theta} 
		\left[ m_0(H_0) 
		-2m_0\bigl(H_0(1-\cos\theta)/2\bigr) \right] 
\nonumber\\
\label{eq_chi-m0}\\
	&=& \frac{4}{in\pi H_0} \int_0^{H_0} dh 
		\exp\left[ in\arccos\left(1-\frac{2h}{H_0}\right) \right] 
		\frac{\partial m_0(h)}{\partial h} . 
\nonumber
\end{eqnarray}
When $H_0\to 0$, all components of the ac susceptibility vanish except for $\chi_1'= (\partial m_0/\partial H_0)_{H_0\to 0}$. 
The ac loss of a SC/FM strip per unit length, $Q$, in an ac magnetic field is proportional to $\chi_1''$ as~\cite{Clem94} 
\begin{equation}
	Q(H_0)= \pi\mu_0 H_0^2 \chi_1''(H_0) . 
\label{eq_Q-X''}
\end{equation}

\begin{figure}[bt]
	\includegraphics{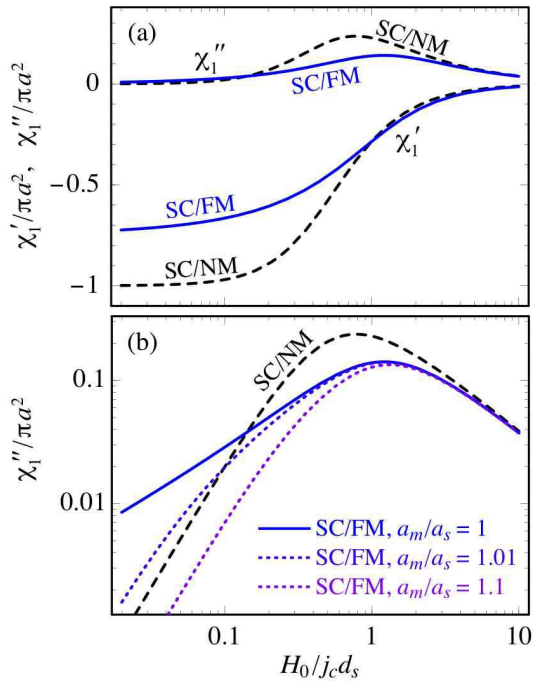}
\caption{(Color onine) %
ac susceptibility $\chi_1'+i\chi_1''$ (in units of $\pi a^2$) for the fundamental frequency ($n=1$) as a function of the amplitude of an applied ac magnetic field, $H_0$, (in units of $j_cd_s$) for a SC/FM strip (solid lines) and for a SC/NM strip (dashed lines).  
(a) Semi-log plot of the real part $\chi_1'$ (lower lines) and imaginary part $\chi_1''$ (upper lines) vs $H_0$. 
(b) Log-log plot of the imaginary part $\chi_1''$ vs $H_0$ for a SC/FM strip with $a_m/a_s=1$ (solid line) and for SC/FM strips with $a_m/a_s=1.01$ and $1.1$ (dotted lines), where $2a_s$ is the width of the superconducting strip and $2a_m$ is the width of the ferromagnetic substrate. 
}
\label{Fig_susceptibility}
\end{figure}

Figure~\ref{Fig_susceptibility} shows the real part $\chi_1'$ and imaginary part $\chi_1''$ of the ac susceptibility for the fundamental frequency ($n=1$) as a function of $H_0$, calculated by substituting Eqs.~\eqref{eq_H0-beta0} and \eqref{eq_m0-critical} into Eq.~\eqref{eq_chi-m0}. 
Except when the magnetic field is strong ($H_0/j_cd_s\gg 1$), the ac susceptibility of a SC/FM strip (solid lines) is significantly different from that of a SC/NM strip (dashed lines). 
When $H_0/j_cd_s\lesssim 1$, the real part $-\chi_1'$ of a SC/FM strip is smaller than that of a SC/NM strip, and when $H_0/j_cd_s\ll 1$, we have $\chi_1'/\pi a^2=-3/4$, which corresponds to Eq.~\eqref{eq_X0-Hay-Meissner}. 

As shown in Fig.~\ref{Fig_susceptibility}(b), $\chi_1''$ of a SC/FM strip (solid line) is {\em smaller} than that of a SC/NM strip (dashed line) when $H_0/j_cd_s >0.14$, whereas $\chi_1''$ of a SC/FM strip is {\em larger} when $H_0/j_cd_s <0.14$. 
Note that the enhancement of $\chi_1''$ for $H_0/j_cd_s <0.14$ is not due to the ferromagnetic hysteresis in the substrate, because we assume a linear $\bm{B}$-$\bm{H}$ relationship in the ferromagnetic substrate, as described in Sec.~\ref{sec_FM-substrate}. 
When $H_0/j_cd_s\lesssim 1$, the edges of SC/FM strips play crucial roles in $\chi_1''$. 
In addition to the $\chi_1''$ of a SC/FM strip with $a_m=a_s$ (as shown in Fig.~\ref{Fig_SC/FM-strip}), Fig.~\ref{Fig_susceptibility}(b) shows $\chi_1''$ of a SC/FM strip with $a_m>a_s$ (Fig.~\ref{Fig_SC/FM-wide}), where $2a_s$ is the width of the superconducting strip and $2a_m$ is the width of the ferromagnetic substrate. 
(See Appendix~\ref{sec_wide-FM}.) 
Even when the ferromagnetic substrate is only $1\%$ wider than the superconducting strip (i.e., $a_m/a_s=1.01$), $\chi_1''$ is strongly affected when $H_0/j_cd_s\lesssim 0.1$. 
For any $H_0/j_cd_s$, $\chi_1''$ of a SC/FM strip with $a_m/a_s=1.1$ is smaller than that of a SC/NM strip. 
The dependence of $\chi_1''$ on $a_m/a_s$ clearly confirms that the enhancement of $\chi_1''$ of a SC/FM strip with $a_m/a_s=1$ for $H_0/j_cd_s<0.14$ is due to the edge effect of a ferromagnetic substrate.

\section{\label{sec_summary}%
Discussion and Summary} 
The three key theoretical predictions presented in Sec.~\ref{sec_critical-state} for the real part $\chi_1'$ and imaginary part $\chi_1''$ of the ac susceptibility, and the ac loss $Q$ ($\propto H_0^2\chi_1''$) of a SC/FM strip exposed to an ac perpendicular magnetic field $H_{ay}=H_0\cos\omega t$ are as follows. 

(i) The $\chi_1'$ in the weak magnetic field limit (i.e., $\partial m_0/\partial H_0$ for $H_0\ll j_cd_s$) of a SC/FM strip is given by $-(3\pi/4)a^2$, which corresponds to $3/4$ of that of a SC/NM strip. 

(ii) The $\chi_1''$ of a SC/FM strip is larger (smaller) than that of a SC/NM strip in the weak (strong) field regime of $H_0/j_cd_s<0.14$ ($H_0/j_cd_s>0.14$), as evidenced in Fig.~\ref{Fig_susceptibility}(b) by the intersection between the line for $\chi_1''$ vs $H_0$ of a SC/FM strip and that of a SC/NM strip at $H_0/j_cd_s\simeq 0.14$, where the intersection field $\mu_0H_0\simeq 0.14\mu_0j_cd_s$ is on the order of mT for typical coated conductors. 

(iii) When the ferromagnetic substrate is wider than the superconducting strip, $\chi_1''$ of a SC/FM strip for $H_0/j_cd_s\lesssim 1$ is suppressed, and can be smaller than that of a SC/NM strip [Fig.~\ref{Fig_susceptibility}(b)]. 

Suenaga {\em et al}.~\cite{Suenaga} experimentally investigated the effects of ferromagnetic Ni-W alloy tapes on ac losses $Q$ of $\rm YBa_2Cu_3O_7$ coated conductors, and they confirmed that ac losses of ferromagnetic substrates are much smaller than those of superconducting strips. 
(See also Refs.~\onlinecite{Suenaga06} and \onlinecite{Tsukamoto07}.) 
The above theoretical predictions (i) and (ii) agree well with this experimental data. 
The theoretical intersection field $\mu_0H_0\simeq 0.14\mu_0j_cd_s$ of $\chi_1''$ vs $H_0$ (or $Q$ vs $H_0$) described in the prediction (ii) is estimated to be about $4.9\,$mT for $j_c=1.2\times 10^{10}\,{\rm A/m^2}$ and $d_s=2.3\,\mu$m, which agrees well with the experimental data by Suenaga {\it et al}.~\cite{Suenaga} 

Prediction (iii) clearly explains that the enhancement of $\chi_1''$ of a SC/FM strip with $a_m/a_s=1$ when $H_0/j_cd_s<0.14$ is due to the edge effect of a ferromagnetic substrate. 
The edges effects are also seen in Fig.~\ref{Fig_flux-front}(b) (i.e., the intersection of the lines of $b_0$ vs $H_0$) and in Fig.~\ref{Fig_Hy-Kz-x}(a) (i.e., the intersection of $H_y$ vs $x$). 

Although our simple model for ferromagnetic substrates assumes an infinite permeability ($\mu_m/\mu_0\to\infty$) in contrast to $\mu_m/\mu_0\sim 30$ in a Ni-W alloy used in coated conductors,~\cite{Suenaga} the quantitative agreement of our theoretical predictions with the experimental data by Suenaga {\em et al}. suggests that the ideal soft magnet model with an infinite permeability works well when $\mu_0/\mu_m\ll 1$.~\cite{Amemiya07,Suenaga,Genenko99} 

In summary, analytical expressions of the complex field were derived for SC/FM strips. 
The ferromagnetic substrates were regarded as ideal soft magnets with an infinite permeability, and the critical state model was used to calculate the magnetic moment and ac susceptibility of a SC/FM strip exposed to a perpendicular magnetic field. 
The theoretical results of the ac susceptibility of a SC/FM strip exposed to a perpendicular magnetic field agreed well with the experimental data by Suenaga {\it et al}.~\cite{Suenaga}

\section{Acknowledgments} 
I thank M.\ Suenaga for stimulating discussions about his experimental data prior to its publication.

\appendix 
\section{\label{sec_wide-FM}%
A Superconducting Strip on a Wide Ferromagnetic Substrate} 
\begin{figure}[b]
	\includegraphics{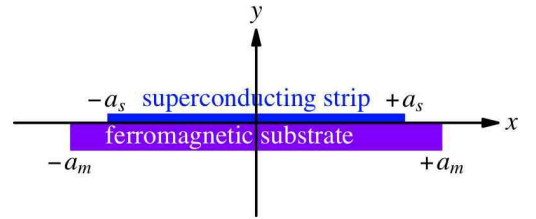}
\caption{(Color onine) %
Cross section of a SC/FM strip with a wider ferromagnetic substrate. 
The superconducting strip is situated at $|x|<a_s$ and $0<y<d_s$, and the ferromagnetic substrate is at $|x|<a_m$ and $-d_m<y<0$, where $d_s+d_m\ll 2a_s< 2a_m$. 
}
\label{Fig_SC/FM-wide}
\end{figure}

The complex field for a SC/FM strip in which the ferromagnetic substrate is wider than the superconducting strip~\cite{Gomory06a,Miyagi07} (i.e., $2a_m>2a_s$ as shown in Fig.~\ref{Fig_SC/FM-wide}) is derived here. 
The SC/FM strip carries no net transport current and is exposed to a perpendicular magnetic field $H_{ay}=H_0$, and the superconducting strip is in the critical state. 

The boundary condition at the surface of the ferromagnetic substrate is given by 
\begin{equation}
	H_x(x,-\epsilon)={\rm Im}\,{\cal H}(x-i\epsilon)=0 
	\quad\mbox{for } |x|<a_m . 
\label{eq_Hx(x,0)=0_wide}
\end{equation}
On the basis of the critical state model, the boundary conditions at the surface of a superconducting strip are 
\begin{eqnarray}
	H_x(x,+\epsilon) &=& {\rm Im}\,{\cal H}(x+i\epsilon)= -{\rm sgn}(x)j_cd_s 
\nonumber\\
	&& \hspace{6em} \mbox{for } b_0<|x|<a_s , 
\label{eq_Hx(x,0)=jcd_wide}\\
	H_y(x,+\epsilon) &=& {\rm Re}\,{\cal H}(x+i\epsilon)=0 
	\quad\mbox{for } |x|<b_0 , 
\label{eq_Hy(x,0)=0_b0_wide}
\end{eqnarray}
where $b_0$ is the parameter for the flux front. 
The corresponding complex field, which satisfies Eqs.~\eqref{eq_Hx(x,0)=0_wide}, \eqref{eq_Hx(x,0)=jcd_wide}, and \eqref{eq_Hy(x,0)=0_b0_wide}, is given by 
\begin{eqnarray}
	\frac{{\cal H}(\zeta)}{2j_cd_s/\pi} &=& 
		{\rm arctanh}\left[ \sqrt{%
		\frac{(\beta_0-\alpha_s)(\eta+a_m)}{%
		(a_m-\alpha_s)(\eta+\beta_0)} } \right] 
\nonumber\\
	&& {}-\frac{\sqrt{(a_m-\alpha_s)(\beta_0-\alpha_s)%
		(\eta+a_m)(\eta+\beta_0)}}{(a_m+\beta_0)\eta} , 
\nonumber\\
\label{eq_H-critical-wide}
\end{eqnarray}
where 
\begin{equation}
	\alpha_s=\sqrt{a_m-a_s} \quad
	\beta_0=\sqrt{a_m-b_0} . 
\label{eq_beta-am-as-b0}
\end{equation}
The parameter $\beta_0$ is related to $H_0$ as 
\begin{equation}
	\frac{H_0}{2j_cd_s/\pi} = 
		{\rm arctanh}\left( \sqrt{\frac{\beta_0-\alpha_s}{a_m-\alpha_s}} \right) 
		-\frac{\sqrt{(a_m-\alpha_s)(\beta_0-\alpha_s)}}{a_m+\beta_0} . 
\label{eq_H0-beta0-wide}
\end{equation}
The magnetic moment per unit length, $m_y=m_0(H_0)$, is 
\begin{equation}
	\frac{m_0}{j_cd_s} = 
		-\left( \frac{2a_m\beta_0}{a_m+\beta_0} +\alpha_s \right) 
		\sqrt{(a_m-\alpha_s)(\beta_0-\alpha_s)} . 
\label{eq_m0-critical-wide}
\end{equation}
The differential susceptibility is simply given by 
\begin{equation}
	\frac{\partial m_0}{\partial H_0} 
	= -\frac{\pi}{4} (3a_m+\beta_0)(a_m-\beta_0) . 
\label{eq_dm0/dH0-wide}
\end{equation}
For $H_0\ll j_cd_s$ (i.e., $\beta_0\simeq \alpha_s$), Eq.~\eqref{eq_dm0/dH0-wide} is reduced to 
\begin{equation}
	\frac{\partial m_0}{\partial H_0} 
	\simeq -\frac{\pi}{4} \left(2a_m^2 +a_s^2 -2a_m\sqrt{a_m^2-a_s^2}\right) , 
\label{eq_dm0/dH0-wide-H0}
\end{equation}
which is further simplified to $\partial m_0/\partial H_0\simeq -(\pi/2)a_s^2$ for $a_m\gg a_s$. 

The ac susceptibility of SC/FM strips with wider ferromagnetic substrates can be calculated by substituting Eqs.~\eqref{eq_H0-beta0-wide} and \eqref{eq_m0-critical-wide} into Eq.~\eqref{eq_chi-m0}. The calculated results for $\chi_1''$ are shown as dotted lines in Fig.~\ref{Fig_susceptibility}(b).

\end{document}